# Planar penta-transition metal phosphide and arsenide as narrow-bandgap semiconductors from first principle calculations

Jun-Hui Yuan,[1,+] Biao Zhang,[1,+] Ya-Qian Song,[1] Jia-Fu Wang,[2] Kan-Hao Xue,[1,3,*] Xiang-Shui Miao[1]

[1] Wuhan National Research Center for Optoelectronics, School of Optical and Electronic Information, Huazhong University of Science and Technology, Wuhan 430074, China

[2] School of Science, Wuhan University of Technology, Wuhan 430070, China

[3] IMEP-LAHC, Grenoble INP – Minatec, 3 Parvis Louis Néel, 38016 Grenoble Cedex 1, France

[+]The authors J.-H. Yuan and B. Zhang contributed equally to this work.

**Corresponding Author**

*E-mail: xkh@hust.edu.cn (K.-H. Xue)

**ABSTRACT**

Searching for materials with single atom-thin as well as planar structure, like graphene and borophene, is one of the most attractive themes in two dimensional materials. Herein, using density functional theory calculations, we have proposed a series of single layer planar penta-transition metal phosphide and arsenide, *i.e.* $TM_2X_4$ (TM= Ni, Pd and Pt; X=P, As). According to the calculated phonon dispersion relation and elastic constants, as well as *ab initio* molecular dynamics simulation results, monolayers of planar penta-$TM_2X_4$ are dynamically, mechanically, and thermally stable. In addition, the band structures calculated with the screened HSE06 hybrid functional including spin-orbit coupling show that these monolayers are direct-gap semiconductors with sizeable band gaps ranging from 0.14 eV to 0.69 eV. Besides, the optical properties in these monolayers are further investigated, where strong in-plane optical absorption with wide spectral range has been revealed. Our results indicate that planar penta-TM$_2$X$_4$ monolayers are interesting narrow gap semiconductors with excellent optical properties, and may find potential applications in photoelectronics.

## 1. Introduction

Since the discovery of atomically thin graphene in 2004, two dimensional (2D) materials have attracted enormous attention in both theoretical and experimental aspects [1–7]. Nowadays, thousands of 2D materials with different compositions and structures have been reported, such as group-IIIA [8–11], group-IVA [12–14], group-VA [3, 15], transition-metal dichalcogenides (TMDCs) [2, 5, 16] and so forth. Among them, 2D phosphorene and phosphides have been intensively studied due to their potential in optoelectronic devices, as well as energy and catalysis applications [17–20]. Recently, a series of phosphides such as GeP [21], GeP$_2$ [22], GeP$_3$ [23], InP$_3$ [24], SnP$_3$ [25–27], CaP$_3$ [28], TlP$_5$ [29] and so forth [30, 31], have been reported to possess excellent electronic and optical properties. For instance, 2D InP$_3$ undergoes a magnetic phase transition under hole doping [24]; single layer SnP$_3$ was found to be a promising anode materials for Na-ion battery [27]; tetragonal-TlP$_5$ shows a direct band gap of 2.02 eV with quite balanced carrier mobilities for electrons (1.396×10$^4$ cm$^2$ V$^{-1}$ s$^{-1}$) and holes (0.756×10$^4$ cm$^2$ V$^{-1}$ s$^{-1}$), even superior to that of phosphorene [29]. Notably, all of these predicted 2D phosphides are of buckling or muti-atomic layer structures.

Very recently, Yuan *et al.* predicted a novel phase of tetragonal PdP$_2$ and PdAs$_2$ with a planar penta-structure [32]. And before that, Liu *et al.* already predicted another planar penta-Pt$_2$N$_4$ based on the crystal prediction software USPEX [33]. Almost simultaneously, we also predicted planar penta-TM$_2$N$_4$ (TM=Ni, Pd and Pt) based on the bulk crystal structure of PtN$_2$ [34]. These works reveal that the transition metals Ni/Pd/Pt prefer to form a planar penta-structures when bonded to group-VA elements. In addition, our previously work also shows the inferior stability of Pd$_2$N$_4$ comparing with Ni$_2$N$_4$ and Pt$_2$N$_4$. Considering the similar chemical properties of Ni/Pd/Pt as well as N/P/As, two questions then arise naturally. (i) Can Ni/Pt and P/As form stable planar penta-structure compounds like Pt$_2$N$_4$ or PdP$_2$? (ii) If so, what are their relative stability and electronic properties comparing with the predicted TM$_2$N$_4$ (TM=Ni, Pd and Pt) and PdP(As)$_2$?

To answer these questions, in this work we have systematically investigated the stability and electronic properties of planar penta-TM$_2$X$_4$ (TM=Ni, Pd and Pt; X=P, As) using first principle calculations. Besides confirming their stability as standalone 2D materials, all planar penta-TM$_2$X$_4$ (TM=Ni, Pd and Pt; X=P, As) are found to be direct band gap semiconductors with energy band gap values ranging from 0.14 eV to 0.77 eV. Photo-absorption spectra have been analyzed with particular attention, where the results support strong optical absorption of these monolayers, especially in the infrared and visible light regimes.

## 2. Computational methods

All calculations were carried out using the plane-wave-based Vienna *Ab initio* Simulation Package (VASP) [35, 36], with a fixed 500 eV plane-wave kinetic energy cutoff. Generalized gradient approximation (GGA) was used for the exchange-correlation (XC) energy, within the Perdew–Burke–Ernzerhof (PBE) functional form [37]. In addition, the screened HSE06 hybrid functional [38] has been used to calculate more accurate band structures in order to overcome the band gap problem of GGA-PBE. The electrons considered as in the valence were: 3$d$ and 4$s$ for Ni; 4$d$ and 5$s$ for Pd; 5$d$ and 6$s$ for Pt; 3$s$ and 3$p$ for P; 4$s$ and 4$p$ for As. Core electrons were approximated by projector augmented-wave pseudopotentials [39, 40]. In all self-consistent runs, the convergence criterion for total energy was set to $10^{-6}$ eV. Moreover, structural optimization was obtained until the Hellmann-Feynman force acting on each atom was less than 0.01 eV/Å in any direction. A 10 × 10 × 1 Monkhorst-Pack $k$-grid for Brillouin zone sampling was used for geometry optimization, which was enlarged to 16 × 16 × 1 for static total energy calculations [41]. In order to reduce the interaction between neighboring layers, a vacuum slab of 20 Å along the $z$-axis was introduced for the 2D monolayers. The phonon dispersion was calculated with the density functional perturbation theory, using the PHONOPY code [42]. We also performed *ab initio* molecular dynamics (AIMD) simulations to evaluate the thermal stabilities. In the AIMD simulations, the PBE functional and NVT canonical ensemble were used, and a 4×4×1 supercell was annealed at various temperatures. Each simulation lasted 5 *ps* with a time step of 1.0 *fs*.

## 3. Results and discussion

### *3.1 Crystal structure and stability*

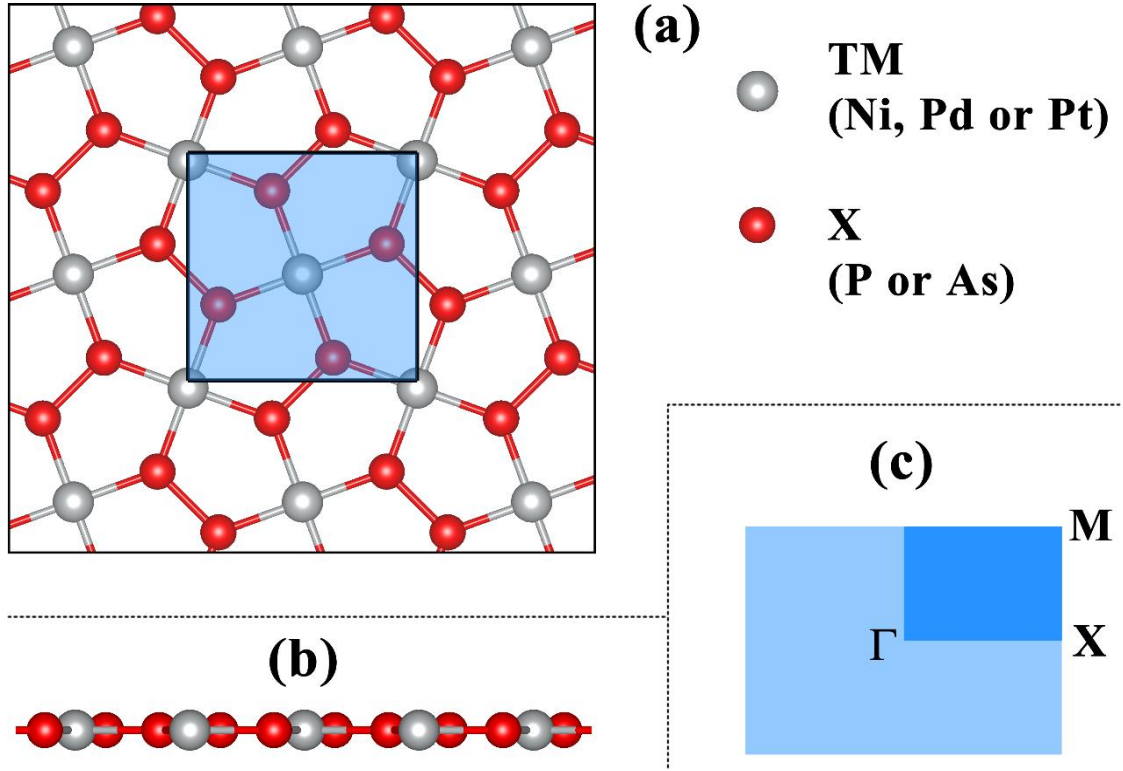

**Figure 1** Crystal structures of pp-TM$_2$X$_4$ (TM=Ni, Pd and Pt; X=P, As): (a) top view, and (b) side view; (c) The 2D Brillouin zone of the monolayers, with high symmetry points marked. Light blue rectangles delimit the unit cell.

**Figure 1** shows the predicted crystal structures of planar penta-TM$_2$X$_4$ (TM=Ni, Pd and Pt; X=P, As), for which we shall use the notation pp-TM$_2$X$_4$ for short. Over the six pp-TM$_2$X$_4$ monolayers, the metal atom always resides at the center of a square planar crystal field formed by four phosphorus or arsenic dimers, leading to four pentagon rings in a unit cell, as demonstrated in **Fig. 1(a)**. The calculated structural parameters are listed in **Table 1**, where the optimized lattice constants are 5.55/5.86/5.83 Å for pp-(Ni/Pd/Pt)$_2$P$_4$, respectively. The corresponding bond lengths (*i.e.* $l_{TM-P}$ and $l_{P-P}$) are 2.16/2.11 Å for pp-Ni$_2$P$_4$, 2.32/2.06 Å for pp-Pd$_2$P$_4$, and 2.30/2.07 Å for pp-Pt$_2$P$_4$, respectively. The P-P bond distances of pp-TM$_2$P$_4$ are smaller than that of black phosphorus ($l_{P-P}$=2.42/2.38 Å) [43] and blue phosphorus ($l_{P-P}$=2.27 Å) [44], indicating the stronger P-P bonding in pp-TM$_2$P$_4$. For pp-TM$_2$As$_4$, the lattice constants are 5.89/6.19/6.16 Å for pp-(Ni/Pd/Pt)$_2$As$_4$, and the corresponding bond lengths (*i.e.* $l_{TM-As}$ and $l_{As-As}$) are 2.27/2.34 Å for pp-Ni$_2$As$_4$, 2.42/2.31 Å for pp-Pd$_2$As$_4$, and 2.40/2.32 Å for pp-Pt$_2$As$_4$, respectively. The As-As bond lengths in pp-TM$_2$As$_4$ are smaller than that of buckled arsenene ($l_{As-As}$=2.486 Å) [45] and puckered arsenene ($l_{As-As}$=2.501/2.485 Å) [46], suggesting the strong As-As bonding in pp-TM$_2$As$_4$, similar to that of pp-TM$_2$P$_4$. The anomalous change in lattice constants and bond lengths of pp-Pd$_2$X$_4$ (X=P, As) have been noticed, which was also observerd in pp-TM$_2$N$_4$ (TM=Ni, Pd and Pt) (**Table 1**) [34]. Since the bond length is related to the bonding strength and the nature of chemical bonds, shorter bond lengths reflect stronger bonding and higher stability. Therefore, it is inferred that the anomalous behavior in pp-Pd$_2$X$_4$ (X=P, As) will directly affect the relative stability of its structure.

**Table 1** Calculated lattice constant *a*, bond lengths $l_{TM-X}$, $l_{X-X}$, cohesive energy $E_c$ and the amount of charge transfer $T_e$ from the TM atom to the X atom (in Bader gauge) of pp-TM$_2$X$_4$ monolayers. The results from Ref. [32] have also been listed for comparison.

|  | *a* (Å) | $l_{TM-X}$ (Å) | $l_{X-X}$ (Å) | $E_c$ (eV) | $T_e$ (e) TM | X |
|---|---|---|---|---|---|---|
| pp-Ni$_2$P$_4$ | 5.55 | 2.16 | 2.11 | 4.09 | +0.357 | -0.160/0.197 |
| pp-Pd$_2$P$_4$ | 5.86 | 2.32 | 2.06 | 3.67 | +0.107 | -0.027/0.080 |
| [32] | 5.86 | 2.32 | 2.06 | 3.67 | -- |  |
| pp-Pt$_2$P$_4$ | 5.83 | 2.30 | 2.07 | 4.49 | -0.064 | +0.006/0.058 |
| pp-Ni$_2$As$_4$ | 5.89 | 2.27 | 2.34 | 3.55 | +0.131 | -0.051/0.080 |
| pp-Pd$_2$As$_4$ | 6.19 | 2.42 | 2.31 | 3.18 | -0.096 | +0.029/0.067 |
| [32] | 6.19 | 2.42 | 2.31 | 3.18 | -- |  |
| pp-Pt$_2$As$_4$ | 6.16 | 2.40 | 2.32 | 3.90 | -0.311 | +0.133/0.178 |

For 2D material prediction, the stability (including thermal stability, dynamic stability and mechanical stability) is a primary issue to be considered. We first calculated the cohesive energies to evaluate the thermal stability of pp-TM$_2$X$_4$, defined as $E_{coh} = (2E_{TM} + 4E_X - E_{TM_2X_4})/6$, where $E_{TM}$, $E_X$ and $E_{TM2X4}$ are the calculated energies of a single TM atom, a single X atom, and that of pp-TM$_2$X$_4$, respectively. As listed in **Table 1**, the calculated cohesive energies are 4.09/3.67/4.49 eV (per chemical formula TM$_2$X$_4$, the same for below) for pp-(Ni/Pd/Pt)$_2$P$_4$ and 3.55/3.18/3.90 eV for pp-(Ni/Pd/Pt)$_2$As$_4$, respectively. In comparison, the cohesive energies for phosphorene, arsenene and antimonene at the same theoretical level are 3.44 eV [54], 2.989 eV [46] and 2.87 eV [48], respectively. Although our results of pp-Pd$_2$X$_4$ are lower than that of phosphorene, they still show considerable stability comparing with arsenene and antimonene, which have been successfully synthesized nevertheless [3, 49, 50]. Furthermore, the thermal stabilities were also examined by AIMD simulations. The atomic configurations and free energy variations for pp-Ni$_2$P$_4$/Pd$_2$P$_4$/Pt$_2$P$_4$/Ni$_2$As$_4$/Pd$_2$As$_4$/Pt$_2$As$_4$, after heating at 500/500/500/500/300/500 K for 5 *ps* are shown in **Fig. S1**. Neither geometric reconstruction nor bond breaking has been discovered during the whole process, suggesting that these planar monolayers may exist as freestanding 2D materials. The variation of free energy only shows slight oscillation (~0.05 eV/atom). Hence, once the as-studied pp-TM$_2$X$_4$ monolayers are synthesized, it can remain stable and robust at room temperature. We have confirmed the thermodynamic stability of pp-TM$_2$X$_4$ through the cohesive energy analysis and AIMD simulations.

Phonon dispersion calculations were performed to further verify the dynamic stability of pp-TM$_2$X$_4$, and the corresponding phonon spectrum and phonon density of states are given in **Fig. 2**. Although very tiny imaginary frequencies are observed in pp-Ni$_2$P$_4$ (about -6.2 cm$^{-1}$) and pp-Pd$_2$As$_4$ (less than -1.0 cm$^{-1}$), which however stem from the computational error, no imaginary frequencies are observed in other four systems. Therefore, we confirm that all the pp-TM$_2$X$_4$ under investigation demonstrate structure rigidity and stability. In addition, it is noteworthy that the highest frequencies are 544/576/574 cm$^{-1}$ for pp-(Ni/Pd/Pt)$_2$P$_4$, and 332/319/317 cm$^{-1}$ for pp-(Ni/Pd/Pt)$_2$As$_4$, respectively, which are higher or comparable to that of phosphorene (440 cm$^{-1}$) [51] and arsenene (305 cm$^{-1}$) [46]. Besides, the two highest optical modes in pp-TM$_2$P$_4$ originate from the P-P bond, while those of pp-TM$_2$As$_4$ are related to the As-As and TM-As bonds. These phonon dispersion relations and phonon density of states results support that the TM-X and X-X bonds are rather robust in these monolayers.

In addition, we also carried out mechanical stability tests by calculating the independent elastic constants of pp-TM$_2$X$_4$, where the results are listed in **Table S1**. For a mechanically stable 2D sheet, the elastic constants need to satisfy $C_{11}C_{22} - C_{12}^2 > 0$ and $C_{66} > 0$ [52]. As shown in **Table S1**, the tetragonal symmetry of pp-TM$_2$X$_4$ yields $C_{11} = C_{22}$. And all the 2D pp-TM$_2$X$_4$ are mechanically stable since the calculated elastic constants satisfy the conditions mentioned above. On the other hand, the calculated Young's moduli ($E$) are 122.19/105.88/136.77 N·m$^{-1}$ for pp-(Ni/Pd/Pt)$_2$P$_4$ and 97.22/84.29/106.72 N·m$^{-1}$ for pp-(Ni/Pd/Pt)$_2$As$_4$, respectively. Compared to other 2D materials, such as graphene (342.2 N·m$^{-1}$) [53], BN (275.8 N·m$^{-1}$) [54], and penta-graphene (263.8 N·m$^{-1}$) [55], the Young's moduli of pp-TM$_2$X$_4$ are much lower, indicating the smaller stiffness and more susceptible to elastic deformation of the pp-$TM_2X_4$. The calculated Poisson's ratio υ are 0.22/0.30/0.29 for pp-$(Ni/Pd/Pt)_2P_4$ and 0.25/0.34/0.32 for pp-$(Ni/Pd/Pt)_2As_4$, respectively, which are higher than that of graphene (0.173) [53] and BN (0.220) [54]. These six nanosheets are all positive Poisson's ratio materials, unlike negative Poisson's ratio materials such as penta-graphene (-0.068) [55] or phosphorene (-0.027) [43].

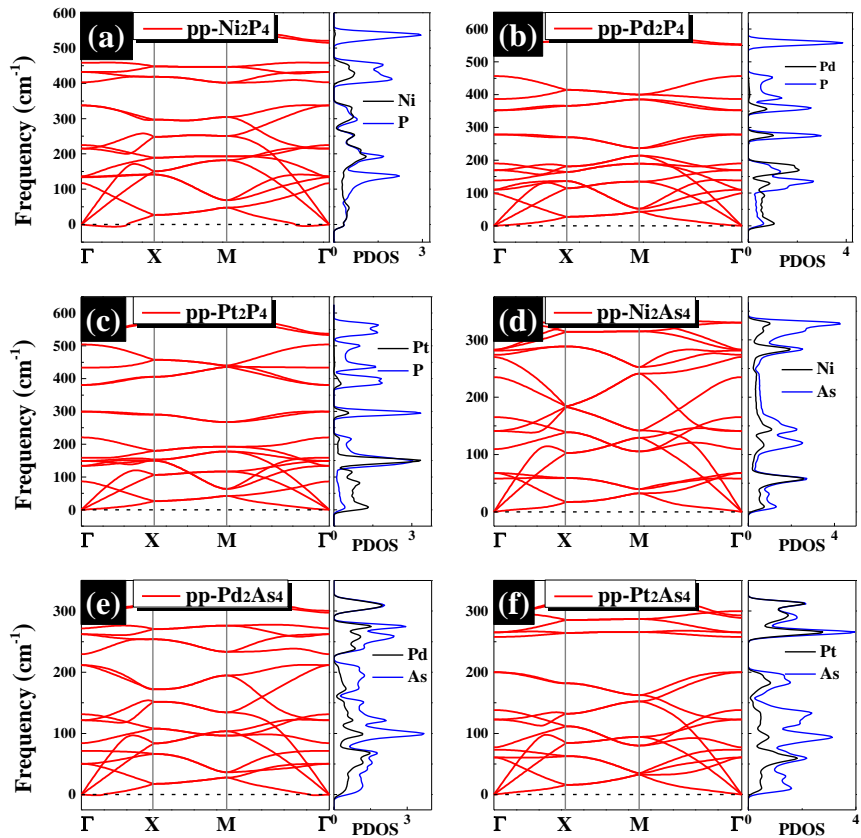

**Figure 2** Phonon spectra and phonon density of states (PDOS) of (a) pp-$Ni_2P_4$; (b) pp-$Pd_2P_4$; (c) pp-$Pt_2P_4$; (d) pp-$Ni_2As_4$; (e) pp-$Pd_2As_4$; and (f) pp-$Pt_2As_4$.

### *3.2 Bonding characteristics*

Next, the electron localization function (ELF) [56–59] and Bader charges [60–62] were calculated to understand the bonding characteristics of pp-$TM_2X_4$. As shown in **Fig. 3**, ELF = 1 represents perfect localization; ELF = 0.5 corresponds to the free electron-gas; and ELF = 0 means the absence of electrons. In pp-TM$_2$X$_4$, most of the electrons are localized in the region of phosphorus (arsenic) atoms, and there are clearly strong covalent bonding between the neighboring phosphorus (arsenic) atoms. Apparently, the P-P bond is stronger than the As-As bond, reflected by the higher value of ELF along the P-P bond. Furthermore, the electrons (at ELF level) around the metallic element gradually increase from Ni to Pt. For instance, there is almost no electrons around Ni in pp-Ni$_2$As$_4$, but an apparently electron distribution around Pt (ELF value of ~0.4) exists in the pp-Pt$_2$As$_4$, which may have significant impact on the electronic structures of pp-TM$_2$X$_4$.

On the other hand, the results of Bader charge analysis are also listed in **Table 1**, for the sake of a deeper understanding into the bonding characteristics of pp-TM$_2$X$_4$. Two distinct scenarios have been observed. In one case, the TM atom lost its valance electrons, which were transferred to the neighboring phosphorus (arsenic) atoms. In the other case, the opposite charge transfer direction was identified. For the first situation, the calculated Bader charge are +0.357 e and {-0.160 e, -0.197 e} for Ni and P in pp-Ni$_2$P$_4$, +0.107 e and {-0.027 e, -0.080 e} for Pd and P in pp-Pd$_2$P$_4$, +0.131 e and {-0.051 e, -0.080 e} for Ni and As in pp-Ni$_2$As$_4$, respectively. For the second situation, the calculated Bader charge are -0.064 e and {+0.006 e, -0.058 e} for Pt and P in pp-Pt$_2$P$_4$, -0.096 e and {+0.029 e, +0.067 e} for Pd and As in pp-Pd$_2$As$_4$, -0.311 e and {+0.133 e, +0.178 e} for Pt and As in pp-Pt$_2$As$_4$, respectively. The anomalous electron transport was consistent with our previous ELF analysis. The two opposite situations can be explained by the difference in electronegativity. The electronegativity of the P and As are 2.19 and 2.18, while those of the TM atoms are 1.91, 2.20 and 2.28 for Ni, Pd and Pt, respectively. The relative order of electronegativity is therefore: Pt>Pd>P>As>Ni. Hence, Ni will lost its valence electrons no matter whether it is bonded to phosphorus or arsenic atoms, while the opposite situation is Pt, which obtains extra electrons from the bonded phosphorus or arsenic atoms. When it comes to Pd, however, both situations may occur as the difference of its electronegativity with respect to phosphorus or arsenic is very tiny.

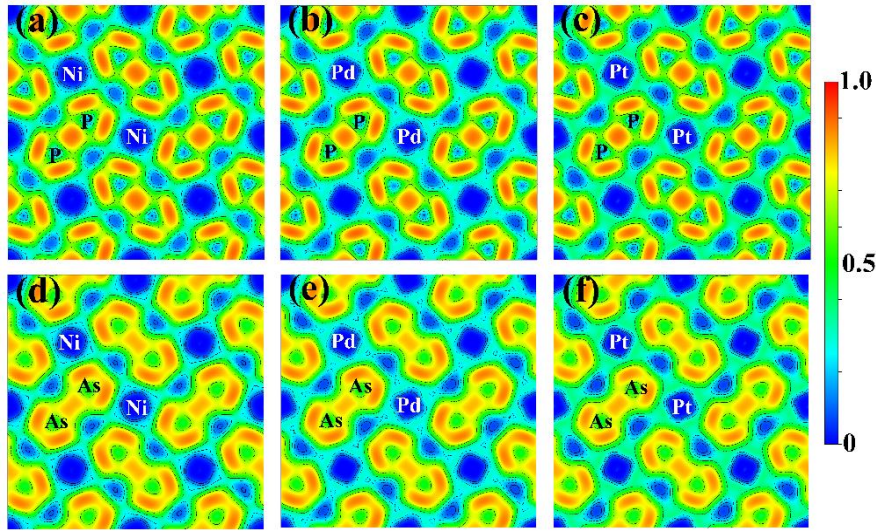

**Figure 3** Electron localization functions (ELFs) of pp-TM$_2$X$_4$ plotted in a 2×2 supercell, calculated with GGA-PBE. (a) pp-Ni$_2$P$_4$; (b) pp-Pd$_2$P$_4$; (c) pp-Pt$_2$P$_4$; (d) pp-Ni$_2$As$_4$; (e) pp-Pd$_2$As$_4$ and (f) pp-Pt$_2$As$_4$.

*3.3 Electronic properties*

Subsequently, we investigated the electronic structures of pp-TM$_2$X$_4$ through calculating their energy band diagrams. Since some heavy elements are involved in pp-TM$_2$X$_4$, the spin-orbit coupling (SOC) effect may not be neglected. Accordingly, the band structures of pp-TM$_2$X$_4$ were calculated either with or without SOC. Our calculated GGA-PBE (GGA-PBE+SOC) band gap values are 0.06(0.09)/0.15(0.16)/0.04(0.08) eV for pp-(Ni/Pd/Pt)$_2$P$_4$ and 0(0)/0.32(0.32)/0.01(0.03) eV for pp-(Ni/Pd/Pt)$_2$As$_4$, respectively (**Table 1**). It is well-known that DFT-GGA tends to severely underestimate the band gaps of semiconductors and insulators [63, 64], thus we further employed the screened HSE06 hybrid functional for more accurate results. According to HSE06 (without and with SOC) calculations, the band gaps of pp-TM$_2$X$_4$ are 0.81(0.65)/0.74(0.69)/0.54(0.37) eV for pp-(Ni/Pd/Pt)$_2$P$_4$ and 0.55(0.37)/0.80(0.43)/0.18(0.14) eV for pp-(Ni/Pd/Pt)$_2$As$_4$, respectively. Thus, expect for pp-Pd$_2$X$_4$ and pp-Pt$_2$As$_4$ whose band gap differences are less than 0.05 eV with and without considering SOC, the SOC has significant influence on the electronic properties of pp-TM$_2$X$_4$ sheets. For example, in pp-Pt$_2$P$_4$ the difference with/without SOC is as large as 0.17 eV. As shown in **Fig. 4**, all pp-TM$_2$X$_4$ are direct band gap semiconductors with the valence band maximum (VBM) and conduction band minimum (CBM) both locating at the *M*-point, similar to that of the penta-Pt$_2$N$_4$ [33, 34]. The band gaps of pp-TM$_2$X$_4$ are in a wide range from 0.14 eV to 0.77 eV, comparable to that of InSb (0.24 eV), InAs (0.41) and Ge (0.74 eV) [65]. The narrow and direct band gap nature imply useful application potential of pp-TM$_2$X$_4$ for solar cells and infrared detectors.

**Table 2** Calculated energy band gaps (PBE and HSE06 with/without spin orbit coupling) of pp-TM$_2$X$_4$ monolayers. The results from Ref. [32] have also been listed for comparison.

|  | PBE (eV) | PBE+SOC(eV) | HSE06 (eV) | HSE06+SOC (eV) |
| --- | --- | --- | --- | --- |
| pp-$Ni_2P_4$ | 0.06 | 0.09 | 0.81 | 0.65 |
| pp-$Pd_2P_4$ | 0.15 | 0.16 | 0.74 | 0.69 |
| [32] | -- | -- | 0.73 | -- |
| pp-$Pt_2P_4$ | 0.04 | 0.08 | 0.54 | 0.37 |
| pp-$Ni_2As_4$ | 0 | 0 | 0.55 | 0.43 |
| pp-$Pd_2As_4$ | 0.32 | 0.32 | 0.80 | 0.77 |
| [32] | -- | -- | 0.81 | -- |
| pp-$Pt_2As_4$ | 0.01 | 0.03 | 0.18 | 0.14 |

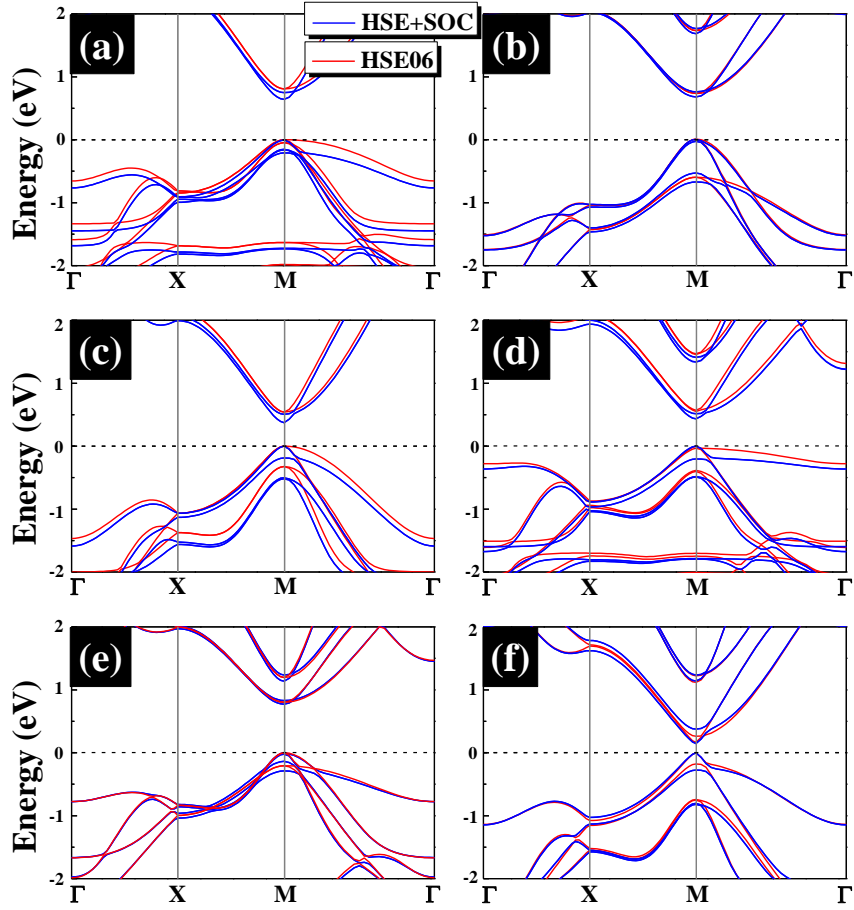

**Figure 4** Electronic band structures of pp-$TM_2X_4$ with and without SOC, calculated with the screened HSE06 hybrid functional. (a) pp-$Ni_2P_4$; (b) pp-$Pd_2P_4$; (c) pp-$Pt_2P_4$; (d) pp-$Ni_2P_4$; (e) pp-$Pd_2P_4$; (f) pp-$Pt_2P_4$.

To further understand the orbital contributions to the electronic structure, we analyzed the total and partial density of states (DOS) of these six monolayers, as shown in **Fig. 5**. The results revealed that the DOS above the Fermi level (0~0.5 eV) for pp-TM$_2$X$_4$ are primarily stemming from P(As)-$p_z$ orbitals. Yet, the DOS below the Fermi level (-0.5 eV~0 eV) is much more complicated. For pp-TM$_2$X$_4$ (TM=Ni, Pt; X=P, As), the P (As)-$p_z$ and Ni (Pt)-$d_{xz}$ orbitals dominate. In the pp-Pd$_2$As$_4$ system, except for the As-$p_z$ and Pd-$d_{xz}$ components, there are also contributions from the As-$s$+$p_x$+$p_y$ orbital. And for pp-Pd$_2$P$_4$, it primarily consists of the P-$s$+$p_x$+$p_y$ orbital. Apparently, there are strong orbital hybridizations between the TM atoms and X atoms, leading to superior stability, just as proven by the previous results of stability analysis.

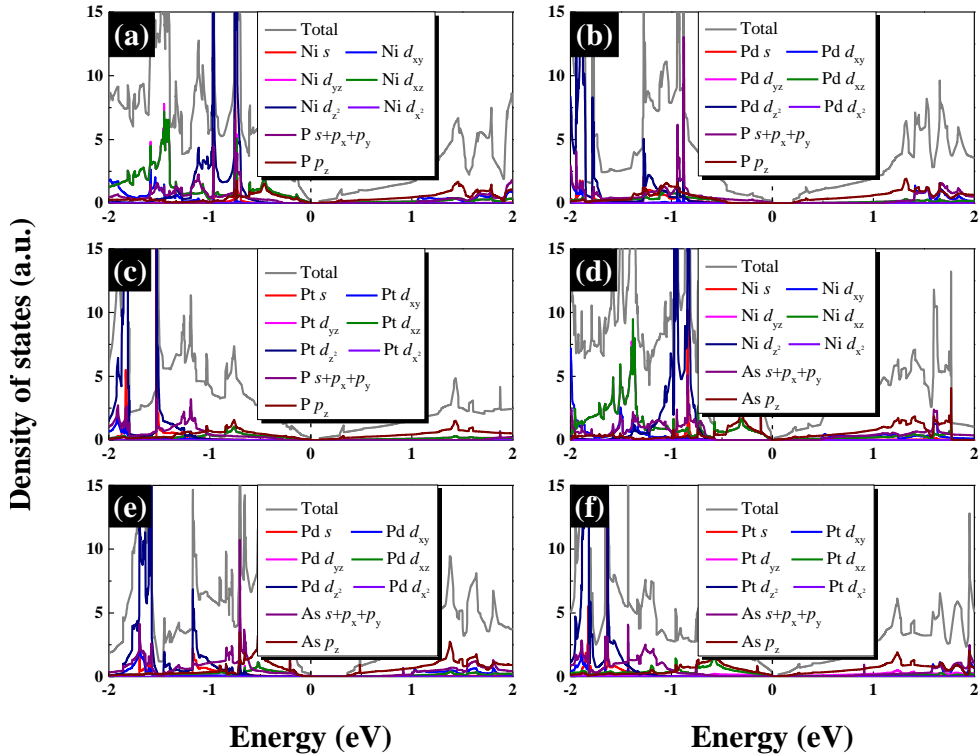

**Figure 5** Density of states of pp-TM$_2$X$_4$ based on GGA-PBE calculation. (a) pp-Ni$_2$P$_4$; (b) pp-Pd$_2$P$_4$; (c) pp-Pt$_2$P$_4$; (d) pp-Ni$_2$P$_4$; (e) pp-Pd$_2$P$_4$; (f) pp-Pt$_2$P$_4$, respectively.

*3.4 Optical properties*

To assess the applicability of pp-TM$_2$X$_4$ monolayers in optoelectronic devices, we calculated their optical properties using the screened HSE06 hybrid functional. The transverse dielectric function $\varepsilon(\omega)$ is used to describe the optical properties of materials [66]: $\varepsilon(\omega) = \varepsilon_1(\omega) + i\varepsilon_2(\omega)$, where $\omega$ is the photon frequency, $\varepsilon_1(\omega)$ and $\varepsilon_2(\omega)$ are the real and imaginary parts of the dielectric function, respectively. The absorption coefficient can be evaluated according to the expression [66] $\alpha(\omega) = \frac{\sqrt{2}\omega}{c}\left\{\left[\varepsilon_1^2(\omega) + \varepsilon_2^2(\omega)\right]^{\frac{1}{2}} - \varepsilon_1(\omega)\right\}^{\frac{1}{2}}$. For comparison, we also calculated the absorption coefficients of silicon at the HSE06 level, as shown in **Fig. 5(a)**. The absorption coefficients of pp-TM$_2$X$_4$ reaches the order of $10^5$ cm$^{-1}$, and covering a wide wave-length range from infra-red (IR), visible light (VL) to the ultraviolet (UV) region. Remarkably, the optical absorptions of pp-TM$_2$X$_4$ in the IR and VL regions are much higher than that of silicon. Its absorption of UV light, however, is weaker than that of silicon. The in-plane absorption is around 10 times that of out-of-plane (**Fig. 5(b)**), due to the larger cross section area. Besides, it has been found that pp-TM$_2$X$_4$ show strongest absorption in the infra-red region.

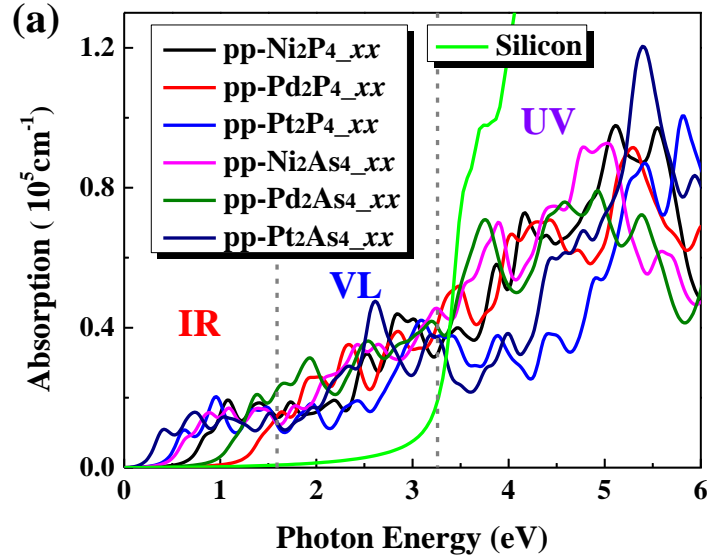

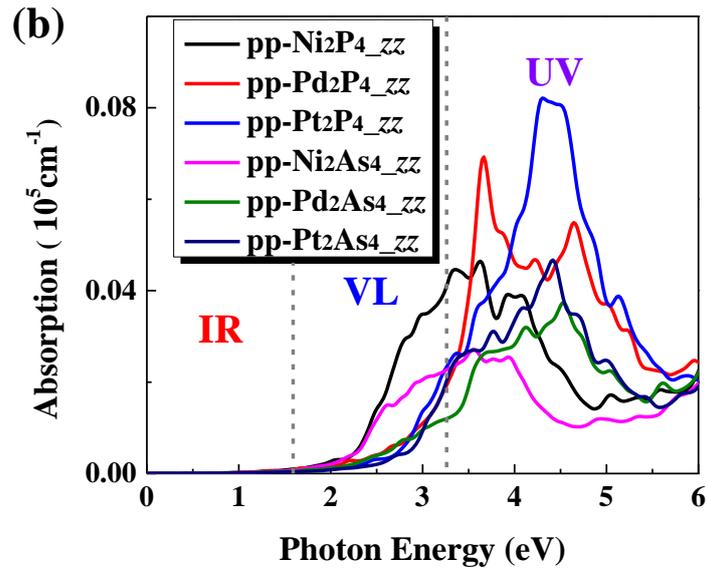

**Figure 6** The (a) in-plane and (b) out-of-plane optical absorption coefficients of pp-$TM_2X_4$ monolayers and silicon, calculated using the screened HSE06 hybrid functional. The energy ranges for the IR (infra-red), VL (visible light) and UV (ultraviolet) regions are separated by vertical lines.

## 4 Conclusions

In summary, a new series of 2D planar penta-$TM_2X_4$ (TM=Ni, Pd and Pt; X=P, As) monolayers have been predicted by first principle calculations. The structures show excellent dynamical, mechanical, and thermal stabilities. Direct band gaps ranging from 0.14 eV to 0.77 eV have been confirmed, considering the effect of spin-orbit coupling. Remarkable optical absorption, especially in the infrared and visible light regions, have been confirmed in the six nanosheets. The pronounced electronic and optical properties may render pp-$TM_2X_4$ monolayers promising materials in solar cells and optoelectronic devices.

**ACKNOWLEDGMENTS**

This work was supported by the National Key Research and Development Program of China (Materials Genome Initiative, 2017YFB0701700), the National Natural Science

Foundation of China under Grant No. 11704134, the Fundamental Research Funds for the Central Universities of China under Grant No. HUST:2016YXMS212, and the Hubei "Chu-Tian Young Scholar" program.

Supporting information for

# Planar penta-transition metal phosphide and arsenide as narrow-bandgap semiconductors from first principle calculations

Jun-Hui Yuan,[1,+] Biao Zhang,[1,+] Ya-Qian Song,[1] Jia-Fu Wang,[2] Kan-Hao Xue,[1,3,*] Xiang-Shui Miao[1]

[1] Wuhan National Research Center for Optoelectronics, School of Optical and Electronic Information, Huazhong University of Science and Technology, Wuhan 430074, China

[2] School of Science, Wuhan University of Technology, Wuhan 430070, China

[3] IMEP-LAHC, Grenoble INP – Minatec, 3 Parvis Louis Néel, 38016 Grenoble Cedex 1, France

[+]The authors J.-H. Yuan and B. Zhang contributed equally to this work.

**Corresponding Author**

*E-mail: xkh@hust.edu.cn (K.-H. Xue)

**Table S1.** Calculated elastic constant $C_{ij}$, Young's modulus ($E$) and Poisson's ratio $\upsilon$ of pp-TM$_2$X$_4$ monolayers.

|  | $C_{11}$ | $C_{22}$ | $C_{12}$ | $C_{66}$ | $E$ (N·m$^{-1}$) | $\upsilon$ |
|---|---|---|---|---|---|---|
| pp-Ni$_2$P$_4$ | 128.12 | 128.12 | 27.57 | 37.40 | 122.19 | 0.22 |
| pp-Pd$_2$P$_4$ | 116.10 | 116.10 | 34.45 | 25.12 | 105.88 | 0.30 |
| pp-Pt$_2$P$_4$ | 149.34 | 149.34 | 43.32 | 38.67 | 136.77 | 0.29 |
| pp-Ni$_2$As$_4$ | 103.88 | 103.88 | 26.30 | 33.04 | 97.22 | 0.25 |
| pp-Pd$_2$As$_4$ | 95.07 | 95.07 | 32.01 | 22.52 | 84.29 | 0.34 |
| pp-Pt$_2$As$_4$ | 119.28 | 119.28 | 38.71 | 32.55 | 106.72 | 0.32 |
| penta-graphene[1] | 265 | 265 | -18 | -- | 263.8 | -0.068 |

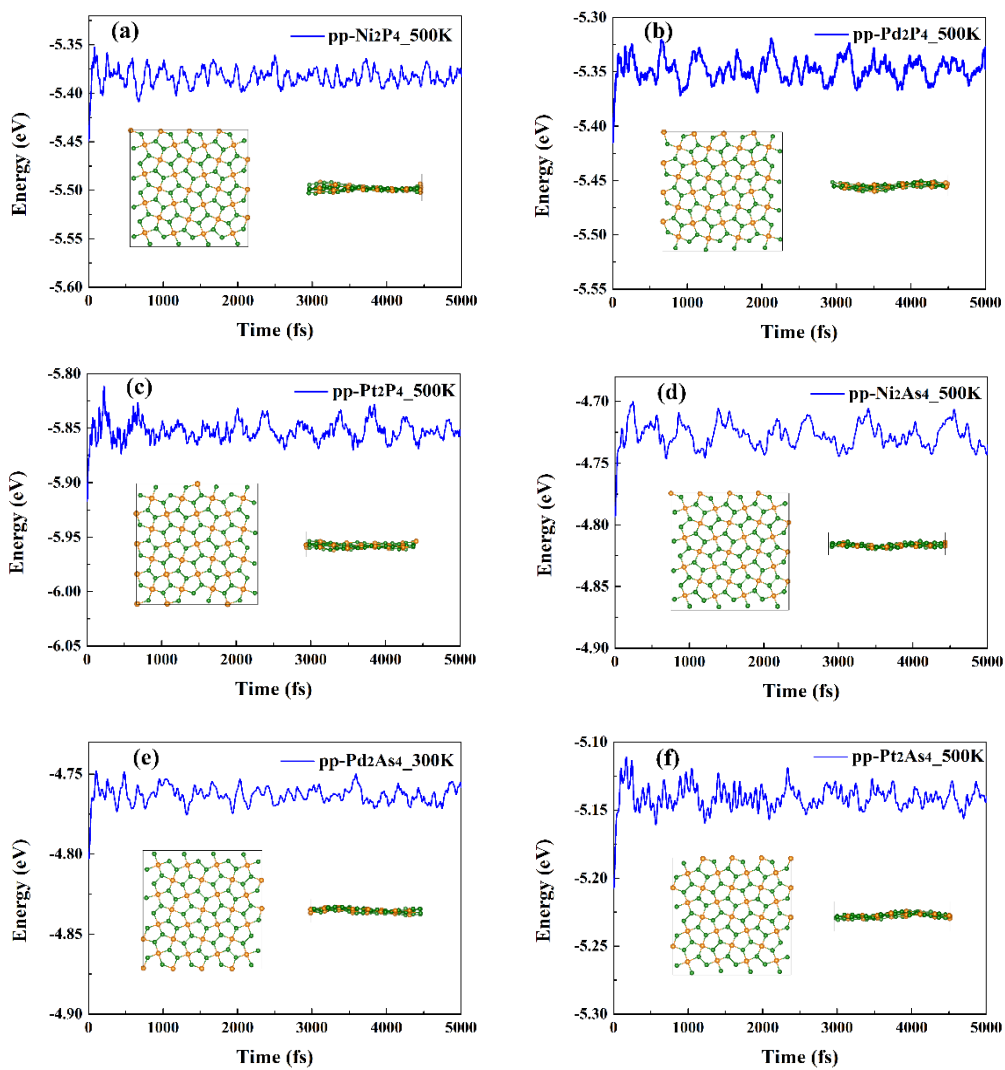

**Figure S1** Variation of the free energy in the molecules dynamic simulation at (a) 500K (for pp-$Ni_2P_4$), (b) 500K (for pp-$Pd_2P_4$), (c) 500K (for pp-$Pt_2P_4$), (d) 500K (for pp-$Ni_2As_4$), (e) 300K (for pp-$Pd_2As_4$) and (f) 500K (for pp-$Pt_2As_4$) during the timescale of 5 *ps*. The insets are the top (right panel) and side (left panel) views of the snapshots from the molecules dynamic simulation of atomic structures for the pp-$TM_2X_4$.

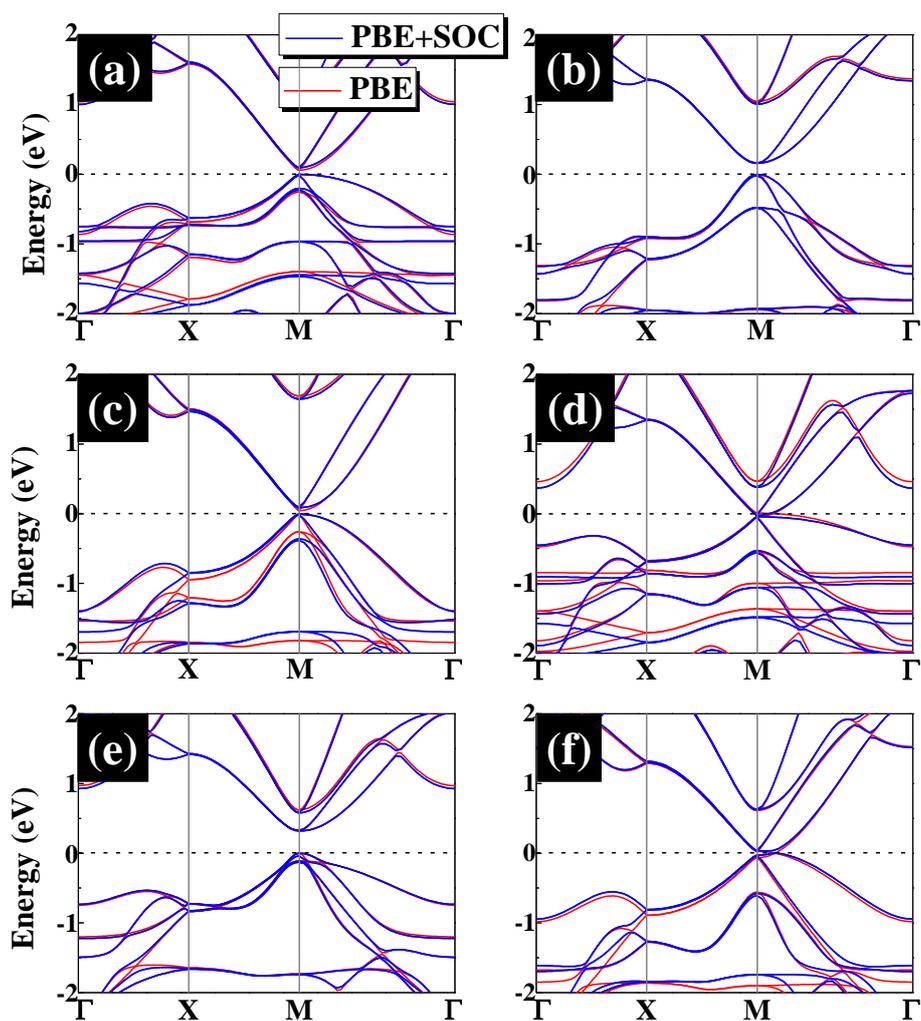

**Figure S2** Band structures of pp-TM$_2$X$_4$ calculated using the PBE functional, either with or without considering the effect of spin orbit coupling (SOC).

**REFRENCES**

[1] S. Zhang, J. Zhou, Q. Wang, X. Chen, Y. Kawazoe, P. Jena, Penta-graphene: A new carbon allotrope, Proceedings of the National Academy of Sciences. 112 (2015) 2372–2377. doi:10.1073/pnas.1416591112.